\documentclass[%
 reprint,https://fr.overleaf.com/project/63d030f841bbca0f21850224
 amsmath,amssymb,
 aps,
]{revtex4-2}

\usepackage{graphicx}
\usepackage{dcolumn}
\usepackage{bm}
\usepackage{xcolor}
\usepackage{siunitx}

\begin{document}

\preprint{APS/123-QED}

\title{Diagnostics of charge breeder electron cyclotron resonance ion source plasma with consecutive transients method}

\author{J. Angot}
\email{julien.angot@lpsc.in2p3.fr}
\affiliation{Univ. Grenoble Alpes, CNRS, Grenoble INP, LPSC-IN2P3, 38000 Grenoble, France}

\author{T. Thuillier}
\affiliation{Univ. Grenoble Alpes, CNRS, Grenoble INP, LPSC-IN2P3, 38000 Grenoble, France}

\author{O. Tarvainen}
\affiliation{UK Science and Technology Facilities Council, ISIS Pulsed Spallation Neutron and Muon Facility, Rutherford Appleton Laboratory, Harwell Campus, OX11 0QX, United Kingdom}

\author{H. Koivisto}
\affiliation{Accelerator Laboratory, Department of Physics, University of Jyv{\"a}skyl{\"a}, FI-40014 Jyv{\"a}skyl{\"a}, Finland}

\author{M. Luntinen}
\affiliation{Accelerator Laboratory, Department of Physics, University of Jyv{\"a}skyl{\"a}, FI-40014 Jyv{\"a}skyl{\"a}, Finland}

\author{V. Toivanen}
\affiliation{Accelerator Laboratory, Department of Physics, University of Jyv{\"a}skyl{\"a}, FI-40014 Jyv{\"a}skyl{\"a}, Finland}





\date{\today}

\begin{abstract}
The consecutive transients (CT) method is a diagnostics approach combining experimental and computational techniques to probe the plasma parameters of Charge Breeder Electron Cyclotron Resonance Ion Sources (CB-ECRIS). The method is based on short pulse injection of singly charged ions into the charge breeder plasma, and the measurement of the resulting transients of the charge bred multiply charged ions. Estimates for plasma density, average electron energy and characteristic times of ion confinement, electron impact ionization and charge exchange are then computationally derived from the experimental data. Here the CT method is applied for parametric studies of CB-ECRIS plasma. Potassium ions were charge bred with hydrogen support plasma, and the effects of varied microwave power, neutral gas pressure and magnetic field strength on the plasma parameters and charge breeding efficiency are presented. It is shown that the method is sufficiently sensitive to provide relevant information on changing plasma conditions with the control parameters. The neutral gas pressure had the strongest impact on the plasma parameters, and the results agree with trends obtained by using other diagnostic methods, e.g. the increase of plasma density with increased neutral gas pressure. Furthermore, the method can provide information inaccessible with other methods, such as the characteristic times of ion confinement, ionization and charge exchange --- and the hierarchy between them. The results show that the peak charge breeding efficiency is obtained for the highest ion charge state for which the ionization time remains shorter than the charge exchange and the ion confinement times.
\end{abstract}

\maketitle


\section{Introduction}

Charge Breeder Electron Cyclotron Resonance Ion Sources (CB-ECRIS) are used in Isotope Separation On-Line (ISOL) -facilities for post-acceleration of radioactive nuclei \cite{Blumenfeld, Wenander}. The charge breeding process involves deceleration and capture of the incident 1+ ion beam, step-wise electron impact ionization to high charge state in the magnetically confined minimum-B ECRIS plasma, and extraction of the charge bred ions together with buffer (or support) gas ions. Optimising the charge breeding efficiency and time benefits from dedicated plasma diagnostics for the CB-ECRIS plasma parameters affecting these steps. 

ECRIS plasmas are unique in many aspects. The electron energy distribution (EED) is strongly non-Maxwellian (see e.g. Ref.~\cite{Izotov_review}) owing to the efficient energy transfer from the microwave electric field to the electrons on a closed magnetic isosurface where the relativistic resonance condition $\omega_{\textrm{RF}}=\omega_{ce}=eB / \gamma m_e$ is met. The resulting electron energies range from a few eV to several hundred keV with the high energy electrons being strongly confined magnetically. The high charge state ions remain relatively cold, i.e. \SI{5}-\SI{30}{\electronvolt} as indicated by the Doppler broadening of their emission lines \cite{Kronholm_Ti}, and are confined electrostatically in a local potential minimum caused by the accumulation of hot electrons in the centre of the trap \cite{Melin, Shirkov}. In fact, simulations \cite{Mascali, Mironov} allude the presence of two (small) potential dips, along the plasma chamber axis, at the mirror points for hot electrons near the ECR-zone. Non-invasive diagnostics methods applied for studying minimum-B ECRIS plasmas (not necessarily charge breeders) include, bremsstrahlung and x-ray diagnostics,  microwave interferometry, plasma diamagnetism measurement, optical emission spectroscopy, measurement of the plasma potential, detection of kinetic instabilities, and escaping electron spectroscopy \cite{Thuillier_review, Mascali_review, Noland, Kronholm_ICIS19, Hitz_review, Tarvainen_plasmapotential, Toivanen_review, Izotov_ICIS19, Tarvainen_ICIS19}.

The consecutive Transient (CT) method \cite{Angot_CT, Luntinen_ICIS21} is a recently developed method combining computational techniques and experiments for probing the plasma density $n_e$, (warm) electron average energy $\left < E_e \right >$, and characteristic times of ion confinement  $\tau_{\textrm{conf}}$, charge exchange  $\tau_{\textrm{cex}}$ and electron impact ionization  $\tau_{\textrm{ion}}$ in CB-ECRIS plasmas. The CT method is based on short pulse 1+ injection into the ECRIS plasma and the analysis of the resulting N+ ion beam transients extracted from the ECRIS. As such, the CT method has both benefits and drawbacks. The positives are: the method can be considered non-invasive, the magnitude of the perturbation it causes can be controlled by adjusting the injected 1+ pulse width and intensity, and the equipment required for the 1+ injection and pulsing as well as N+ current measurement is readily available in all CB-ECRIS facilities. The downsides of the method are: limited combinations of 1+ ions and plasma species (clean charge state distribution spectrum with minimum of 5 consecutive charge states without $m/q$-overlap is necessary for the method), the complexity of data analysis, and the large uncertainties of the characteristic times due to the lack of accurate cross section data for high charge state ionisation. As such, the method either complements existing CB-ECRIS diagnostics while requiring fewer assumptions, or provides information inaccessible through other techniques. For example, it has been shown with the CT method that the ion confinement time is not a linear function but rather increases exponentially with the charge state \cite{Angot_CT, Luntinen_ICIS21}, which is commensurate with electrostatic ion confinement in a local potential dip \cite{Melin, Shirkov}. It has been shown that the inherent uncertainties of the CT method can be reduced e.g. by two-component injection or overlapping the $\left ( \left < E_e \right > , n_e \right )$ solution sets of neighbouring charge states albeit with the caveat of additional assumption \cite{Luntinen_PhysRevE}. Furthermore, it has been demonstrated that the main contributor to the large relative uncertainty of the CT method is the lack of precise ionization cross section data whereas the presumed EED has a smaller effect \cite{Luntinen_uncertainty}. 

In this paper we apply the CT method for parametric studies of the CB-ECRIS plasma. We do not detail the method itself but rather refer the reader to the literature \cite{Angot_CT, Luntinen_ICIS21, Luntinen_PhysRevE, Luntinen_uncertainty} for a comprehensive account of the assumptions, computational details and data analysis.

In the following sections we describe the experimental setup, and present the measured plasma energy content and characteristic times along with the charge breeding efficiency of potassium as a function of the CB-ECRIS microwave power, neutral (hydrogen) pressure and magnetic field strength. These sweeps are carried out to demonstrate that the CT method is sensitive enough to pick up trends in the above plasma parameters (observables) responding to the change of the control parameters. The results of the CT method are placed in context comparing them to the outcomes of other ECRIS plasma diagnostics.   

\section{Experimental setup and procedure}

\begin{figure*}
    \centering
    \includegraphics[width=1.0\textwidth]{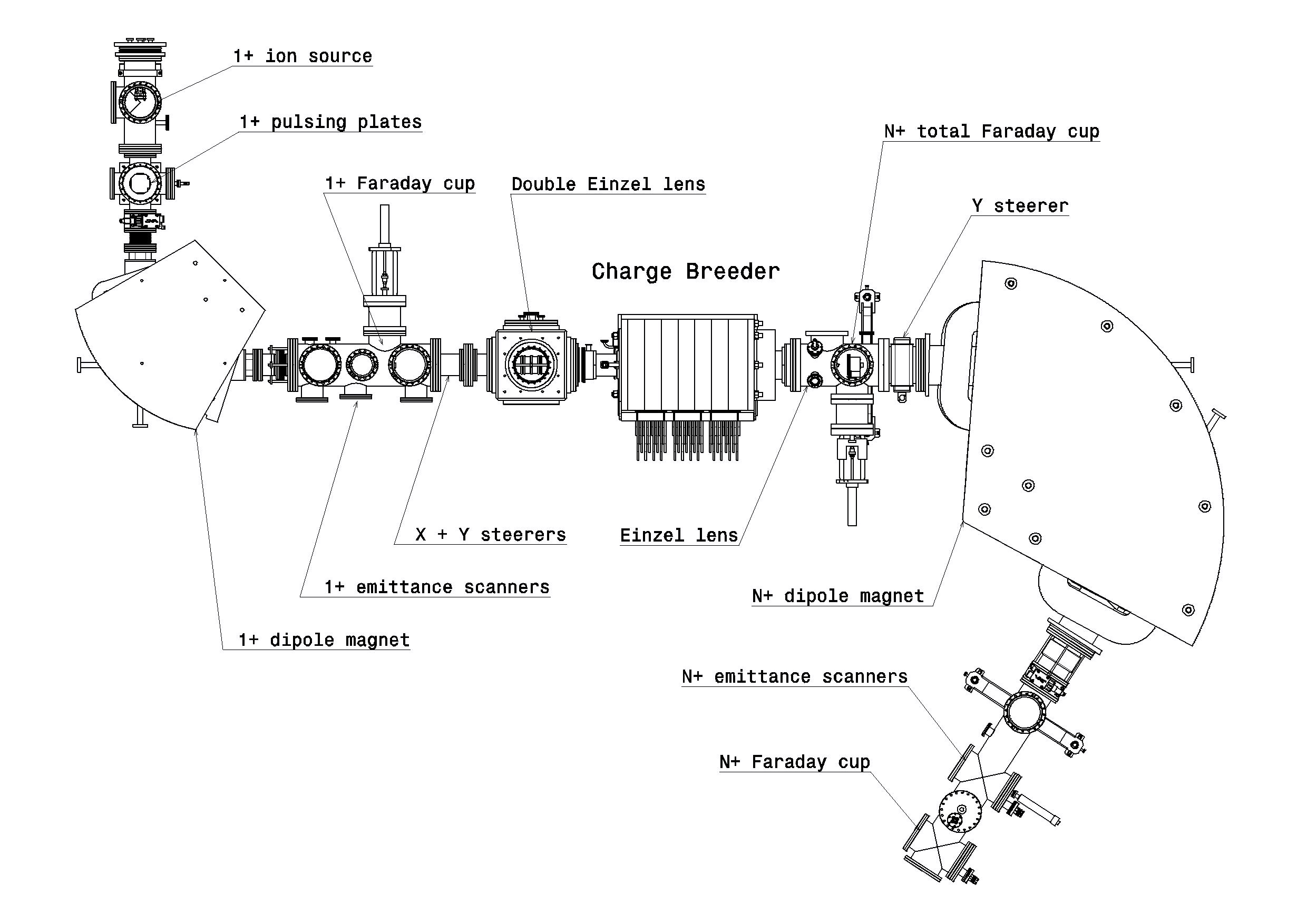}
    \caption{Schematic view of the 1+$\rightarrow$N+ CB-ECRIS test bench.}
    \label{fig:1+N+ test bench}
\end{figure*}

The experiments were carried out on the LPSC 1+$\rightarrow$N+ test bench, shown in Fig.~\ref{fig:1+N+ test bench}, dedicated for the development of the PHOENIX CB-ECRIS \cite{Lamy_2002} --- in particular measurements of the charge breeding efficiency, charge breeding time and $m/q$-contamination. 

Delivering on this remit requires generating a stable 1+ beam with fine tuning of the ion injection energy, a good base vacuum on the order of \SI{e-8}{\milli\bar} or better, the hardware for pulsing of the 1+ beam and beam diagnostics. Hence, the 1+ beam line is equipped with a surface ionisation source producing alkali metal beams, a dipole magnet for mass separation, a Faraday cup to measure the beam intensity, beam optics and deflecting plates for 1+ injection optimisation and pulsing. The 1+ source potential is typically set to HV=\SI{20}{\kilo\volt}. The CB-ECRIS plasma chamber is then biased to HV-$\Delta$V with a negative supply floating at the 1+ source potential. This configuration allows fine-tuning the 1+ ion energy, which is essential for the 1+ beam capture by the CB plasma through electrostatic deceleration by the charge breeder and its plasma potential, and subsequent thermalization of the injected ions in ion-ion collisions with the buffer gas ions. The beams extracted from the charge breeder are analysed in the N+ beam line with a mass spectrometer and diagnostics including a Faraday cup for beam intensity measurement.

The current incarnation of the LPSC charge breeder is a \SI{14.5}{\giga\hertz} minimum-B ECR ion source equipped with three coils to create the axial magnetic profile with two magnetic mirrors at the injection and extraction, respectively \cite{Angot_ECRIS2020}. Typical operational values of the injection, minimum-B and extraction axial magnetic fields are B$_{\textrm{inj}}\approx$~\SI{1.6}{\tesla}, B$_{\textrm{min}}\approx$~\SI{0.4}{\tesla}, and B$_{\textrm{ext}}\approx$~\SI{0.8}{\tesla}. A permanent magnet sextupole surrounding the plasma chamber creates the radial magnetic mirror of \SI{0.8}{\tesla} at the plasma chamber wall, in front of the pole (the total radial field then being affected by the radial component of the solenoid field). A \SI{2}{\kilo\watt} klystron microwave amplifier for plasma (electron) heating is connected to the plasma chamber through a direct waveguide port. The vacuum pumping system assures a base pressure of approximately \SI{3e-8}{\milli\bar} at the source injection. 

Here we apply the CT method to observe the influence of different charge breeder tuning parameters on the plasma characteristics i.e. $n_e$, $\left < E_e \right >$ and $\tau_{\textrm{conf}}$, $\tau_{\textrm{cex}}$ and $\tau_{\textrm{ion}}$ of potassium ($^{39}$K) ions. We chose K as the injected element because it is an alkali element, thus minimising the wall recycling, with several (consecutive) charge states from K$^+$ to K$^{12+}$ found in the $m/q$ spectrum without overlap with the support or residual gas ions. Hydrogen was chosen as plasma support gas to obtain high charge breeding efficiencies of high charge state K ions. The ion source control parameters varied systematically in this study were: (i) the microwave power as it presumably influences the EED and plasma density (see e.g. Refs.~\cite{Noland, Sakildien, Ropponen}), (ii) the support gas feed rate (pressure) which acts on the neutral and electron densities (see e.g. Refs.~\cite{Izotov_LEED, Sakildien}), (iii) the magnetic field minimum B$_{\textrm{min}}$ as it affects the tail of the EED and the occurrence of kinetic plasma instabilities \cite{Izotov_LEED2, Bhaskar, Tarvainen_instability}, and (iv) the extraction magnetic field B$_{\textrm{ext}}$, which allegedly affects the trapping of the hot electrons and the global plasma confinement \cite{Toivanen_ganil}. For each parameter sweep, a \SI{500}-\SI{900}{\nano\ampere} K$^+$ beam was produced with the 1+ ion source. The beam line optics and the CB parameters were optimized for the charge breeding of K$^{9+}$ resulting in 19-20\% efficiency at best. The 1+ pulsing was then used to generate \SI{10}{\milli\second} bunches of injected ions at \SI{1}{\hertz} repetition rate allowing the ion current transients to decay before the onset of the subsequent 1+ pulse. The multi-charged K beam intensity responses (transients) were measured from the Faraday cup of the N+ beam line averaging over 64 waveforms for each charge state to improve the signal-to-noise ratio of the measurement. The microwave power and gas feed rate studies were carried out in the control parameter ranges yielding a stable CB regime i.e. the magnetic field was chosen accordingly to avoid kinetic instabilities. Only one CB-ECRIS control parameter was varied during each sweep. The B$_{\textrm{min}}$ and B$_{\textrm{ext}}$ values corresponding to certain combinations of coil currents were simulated with Radia3D \cite{Radia3D}. The coil currents were then adjusted so that in each sweep only either B$_{\textrm{min}}$ or B$_{\textrm{ext}}$ varied while other field values remained constant. The parameter settings for each sweep are given in Section~\ref{results} along with the data plots. For each setting, the $\Delta$V value was adjusted to optimize the K$^{9+}$ breeding efficiency.

\section{Results}
\label{results}

In the following subsections we present the results of the CB-ECRIS parameter sweeps. We first describe the K$^+ \rightarrow$ K$^{n+}$ charge breeding efficiencies as a function of each parameter. The $(\langle E_e \rangle, n_e)$ solution sets derived from the transients of each ion charge state are used for calculating the plasma energy content $n_e\left < E_e \right >$, which is then presented along with the characteristic times.

\begin{figure}
\includegraphics[width=\columnwidth]{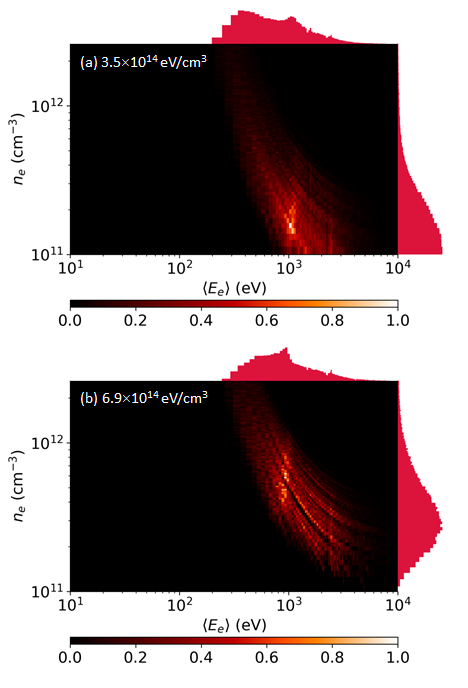}
\caption{\label{fig:solution_set_example} Examples of the $\left ( \left < E_e \right > , n_e \right )$ solution sets for K$^{10+}$ and the corresponding plasma energy contents. The H$_2$ gas pressure was increased from \SI{6.8e-8}{\milli\bar} in (a) to \SI{2.3e-7}{\milli\bar} in (b). The energy content increases due to increased median plasma density from \SI{3.8e11}{\per\centi\meter\cubed} to \SI{1.5e12}{\per\centi\meter\cubed}.}
\end{figure}

For clarity, we present two examples of the $\left ( \left < E_e \right > , n_e \right )$ solution sets for K$^{10+}$ in Figs.~\ref{fig:solution_set_example}(a) and \ref{fig:solution_set_example}(b) highlighting the change of the calculated plasma energy content from $3.5\times 10^{14}$\SI{}{\electronvolt/\centi\meter\cubed} to $6.9\times 10^{14}$\SI{}{\electronvolt/\centi\meter\cubed} with the notable shift of the $\left ( \left < E_e \right > , n_e \right )$ solution space towards higher plasma density. The energy content value is taken as the median value of the product of the $\left ( \left < E_e \right > , n_e \right )$ solutions from the CT-method. These examples were measured as a part of the gas pressure sweep discussed later. The $n_e$ and $\left < E_e \right >$ values were restricted to \SI{e11}{\per\centi\meter\cubed} $\leq n_e \leq$ \SI{2.6e12}{\per\centi\meter\cubed} 
and \SI{10}{\electronvolt} $\leq E_e \leq$ \SI{10}{\kilo\electronvolt}. These limits are based on experimental evidence and simulations of the electron density, and electron energy as explained in Ref.~\cite{Angot_CT} and references therein.

The plasma energy contents and characteristic times are presented in Sections \ref{sec:uw} - \ref{sec:bext} without the estimated uncertainties for the sake of illustration clarity. The uncertainties are discussed separately in Section \ref{sec:uncertainty}. 

\subsection{Microwave power}\label{sec:uw} 

\begin{figure}
\includegraphics[width=0.5\textwidth]{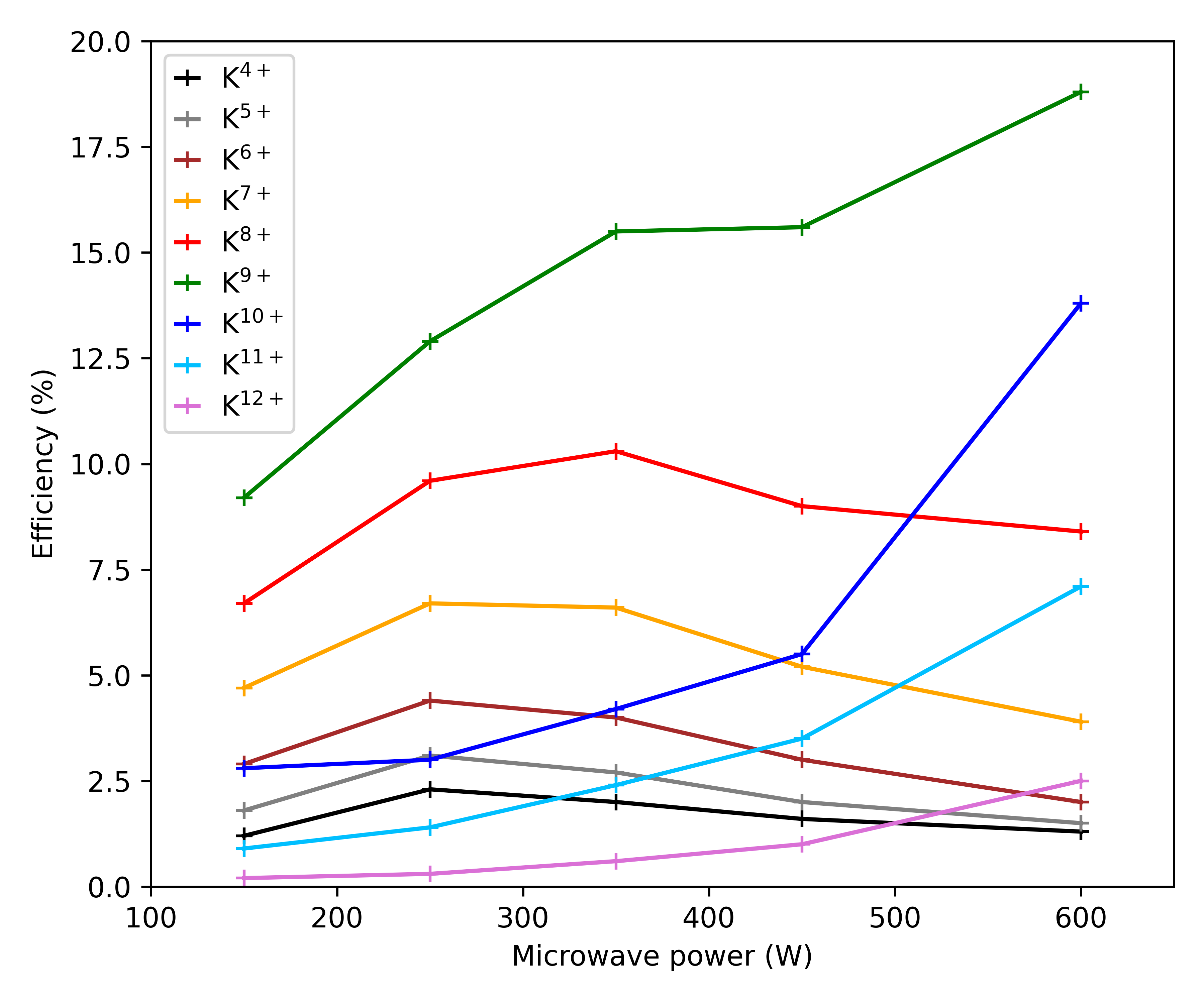}
\caption{\label{fig:uw_efficiency} The charge breeding efficiency of K$^{4+}$ - K$^{12+}$ as a function of the microwave power.}
\end{figure}

The charge breeding efficiencies of K$^{4+}$-K$^{12+}$ as a function of the microwave power are shown in Fig.~\ref{fig:uw_efficiency}. Increasing the power increases the average charge state of K ions, which causes the breeding efficiency of K$^{9+}$-K$^{11+}$ to improve significantly with this control parameter. In contrast, the efficiency of charge states $\leq$K$^{8+}$ exhibit a maximum efficiency at \SI{350}{\watt} and then a decrease as the power is ramped up. Two sweeps were made to ensure the reproducibility of the observed trends but are only shown for the plasma energy content plots. The increase of the high charge state breeding efficiency with the microwave power was observed in both microwave power sweeps.  The other source settings in these sweeps were as follows: neutral gas pressure \SI{1.2e-7}-\SI{1.3e-7}{\milli\bar}, B$_{\textrm{inj}}$ \SI{1.57}-\SI{1.58}{\tesla}, B$_{\textrm{min}}$ \SI{0.44}-\SI{0.45}{\tesla} and B$_{\textrm{ext}}$ \SI{0.84}{\tesla}.

\begin{figure}
\includegraphics[width=0.5\textwidth]{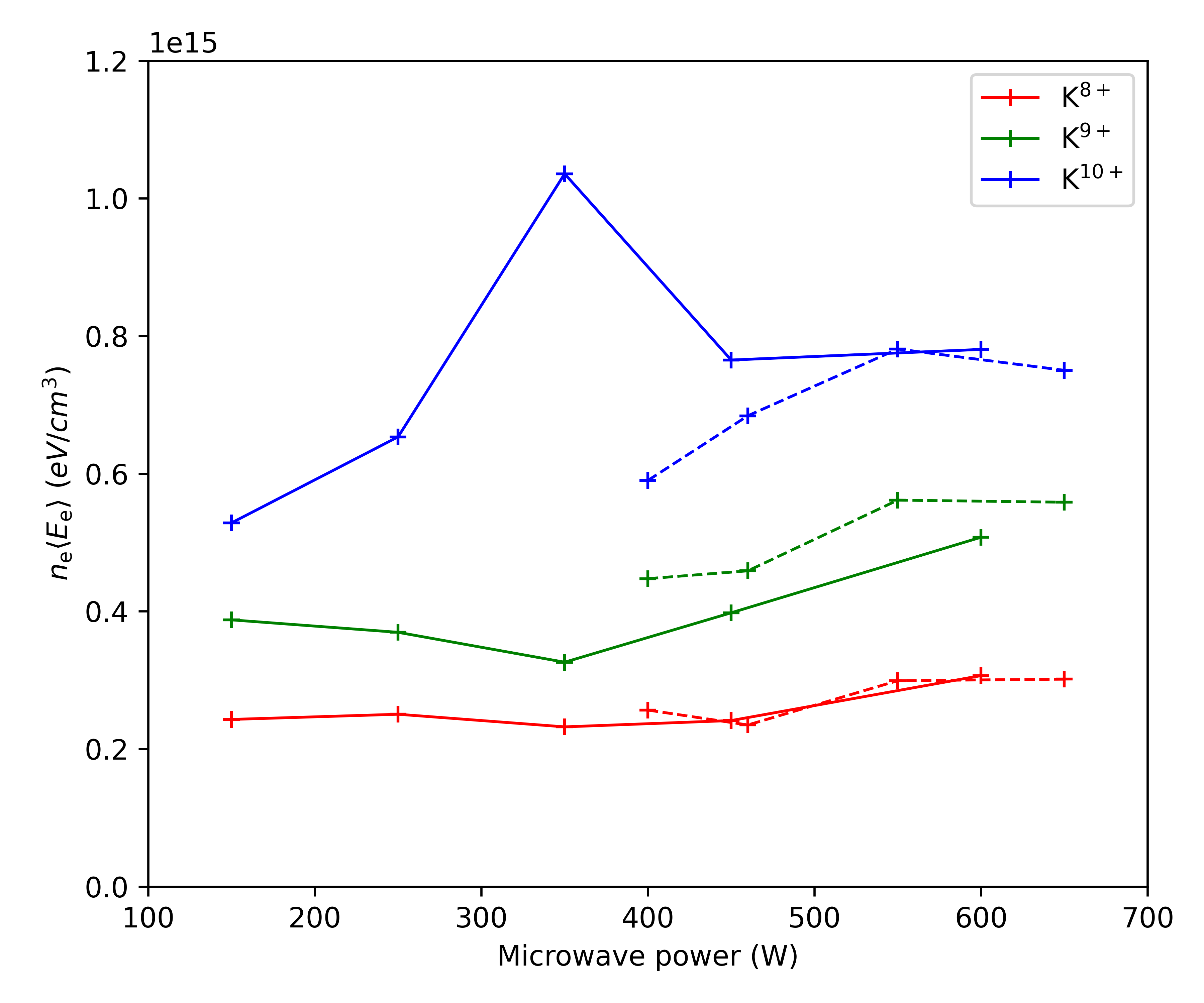}
\caption{\label{fig:uw12_energycontent} The (local) plasma energy content of K$^{8+}$, K$^{9+}$ and K$^{10+}$ as a function of the microwave power. The solid and dashed lines represent two different power sweeps.}
\end{figure}

Figure~\ref{fig:uw12_energycontent} shows the (median) energy content as a function of the microwave power for potassium charge states from K$^{8+}$ to K$^{10+}$ in the two microwave power sweeps. Three observations can be made; (i) the calculated energy content depends on the charge state, which is attributed to the highest charge states originating from the core of the plasma where the plasma density and electron energies can be argued to be higher, (ii) the trend of the plasma energy content is to increase with microwave power by 20-40\% (ignoring a single outlier data point for K$^{10+}$ at \SI{350}{\watt}), and (iii) the trend of the energy content was found to be similar for both microwave power sweeps.

\begin{figure*}
\includegraphics[width=\textwidth]{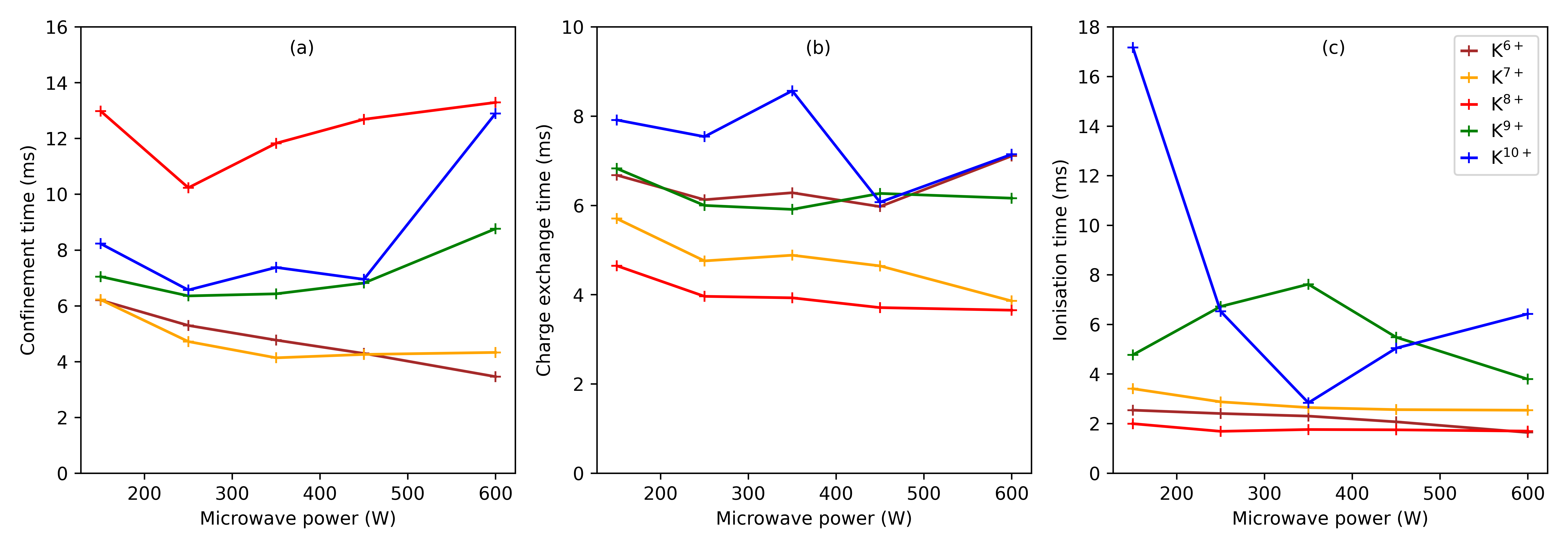}
\caption{\label{fig:uw_time}  The confinement, charge exchange and ionisation times ($\tau_{\textrm{conf}}$, $\tau_{\textrm{cex}}$ and $\tau_{\textrm{ion}}$) of K$^{6+}$ - K$^{10+}$ ions as a function of the microwave power.}
\end{figure*}

Figure \ref{fig:uw_time} shows $\tau_{\textrm{conf}}$, $\tau_{\textrm{cex}}$ and $\tau_{\textrm{ion}}$ for charge states K$^{6+}$-K$^{10+}$ as a function of the microwave power. The confinement time of the high charge states K$^{8+}$-K$^{10+}$ is longer than the confinement time of the charge states K$^{6+}$-K$^{7+}$, which is commensurate with the spatial distribution of the ions found in simulations and experiments \cite{Mironov, Panitzsch, Emilie} suggesting that the highest charge states are highly collisional and electrostatically confined (as opposed to magnetically confined electrons). The trend of $\tau_{\textrm{conf}}$ with the microwave power is to decrease for charge states K$^{6+}$-K$^{7+}$ and to increase for charge states K$^{8+}$-K$^{10+}$. Admittedly there are data points deviating from the trend, and altogether the variation of $\tau_{\textrm{conf}}$ is not drastic. Nevertheless, we draw the attention to the fact that the confinement time of the highest charge states, i,e. K$^{8+}$-K$^{10+}$, are always longer than their ionization times. Furthermore, in the case of K$^{10+}$ the notable increase of $\tau_{\textrm{conf}}$ at high microwave power is correlated with significant increase of the corresponding charge breeding efficiency. The charge exchange time is found longest for the high charge state ions and the trend of $\tau_{\textrm{cex}}$ is to decrease with the microwave power, but again, the changes are very small. Finally, $\tau_{\textrm{ion}}$ does not exhibit a clear trend with the microwave power, but is shorter for the lower charge states than for the higher ones. This is probably due to very low neutral gas density in the core plasma (potential dip) where the highest charge states reside as discussed in Ref.~\cite{Mironov}. The ionization rate coefficient first increases and then typically plateaus towards high $\langle E_e \rangle$ (e.g. $>$ \SI{1000}{\electronvolt} for K$^{8+}$), so improved electron heating is not expected to affect $\tau_{\textrm{ion}}$ by much assuming sufficiently high $\langle E_e \rangle$ as indicated by the solution sets. 
Generally speaking, high charge state production requires $\tau_{\textrm{ion}}$ to be shorter than $\tau_{\textrm{cex}}$, which is the case found throughout the power sweep for most of the charge states, especially those below the K$^{9+}$ peak charge state of the CB-efficiency. Finally, we note that the continuous increase of the K$^{9+}$ efficiency with the microwave power is accompanied with a decrease of $\tau_{\text{ion}}$ of K$^{6+/7+}$ and an increase of $\tau_{\text{conf}}$ of the higher charge state ions. This highlights the importance of the hierarchy of the characteristic times in regards to the optimum charge state for the highest CB efficiency.



\subsection{Neutral gas pressure}

\begin{figure}
\includegraphics[width=0.5\textwidth]{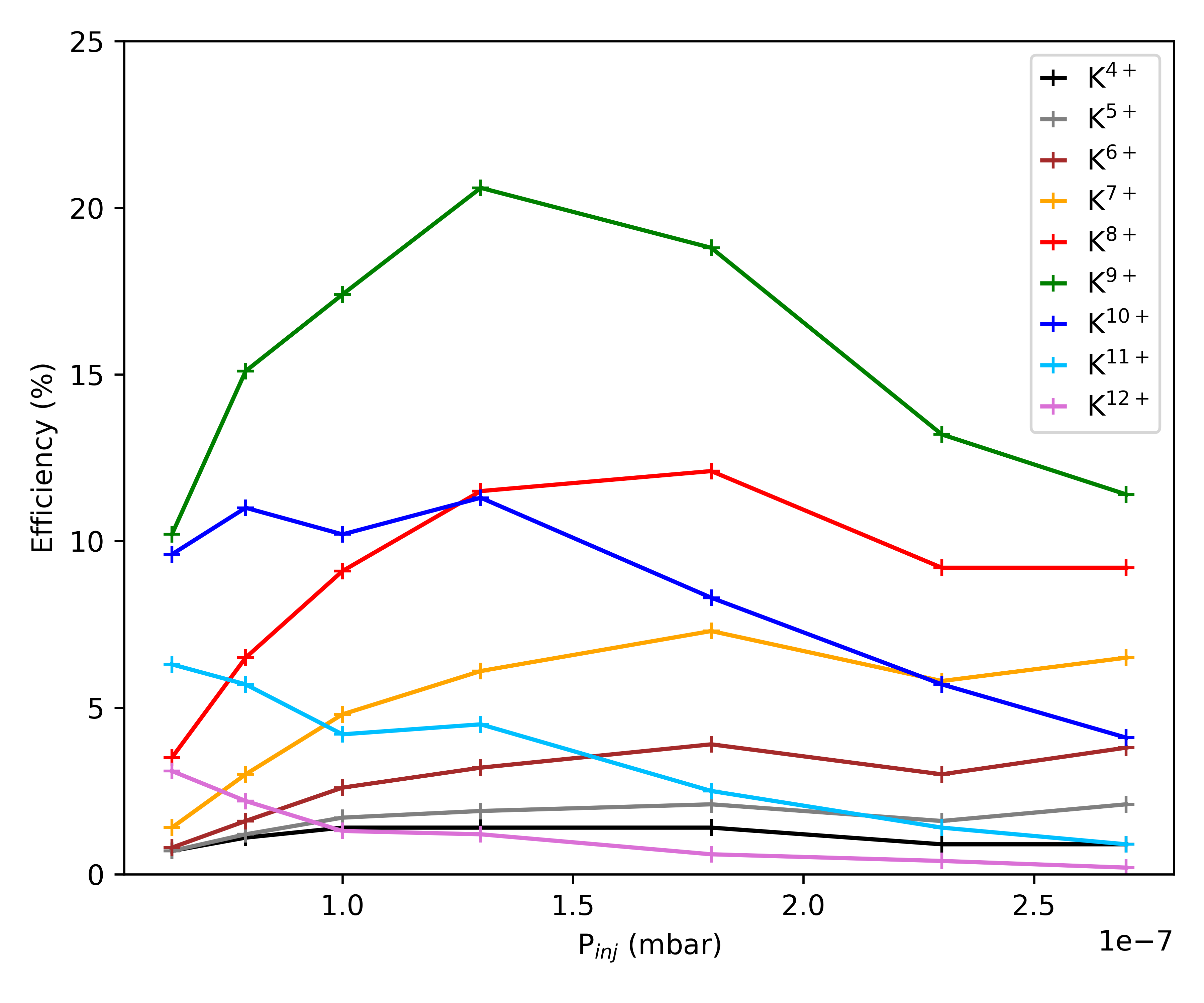}
\caption{\label{fig:gas_Efficiency} The charge breeding efficiency of K$^{4+}$ - K$^{12+}$ as a function of the H$_2$ (buffer) gas pressure.}
\end{figure}

The effect of the neutral H$_2$ pressure (gas feed rate) on the charge breeding efficiency and plasma parameters was studied through two sweeps with otherwise almost identical configuration, i.e. \SI{530}-\SI{560}{\watt} microwave power, \SI{1.57}{\tesla} B$_{\textrm{inj}}$, \SI{0.44}-\SI{0.45}{\tesla} B$_{\textrm{min}}$ and \SI{0.83}-\SI{0.84}{\tesla} B$_{\textrm{ext}}$. Fig.~\ref{fig:gas_Efficiency} shows the charge breeding efficiencies of K$^{4+}$-K$^{12+}$ for one of the sweeps. The behaviour is rather complex but can be summarized as follows; the higher the charge state, the lower the optimum pressure for maximising the charge breeding efficiency.

\begin{figure}
\includegraphics[width=0.5\textwidth]{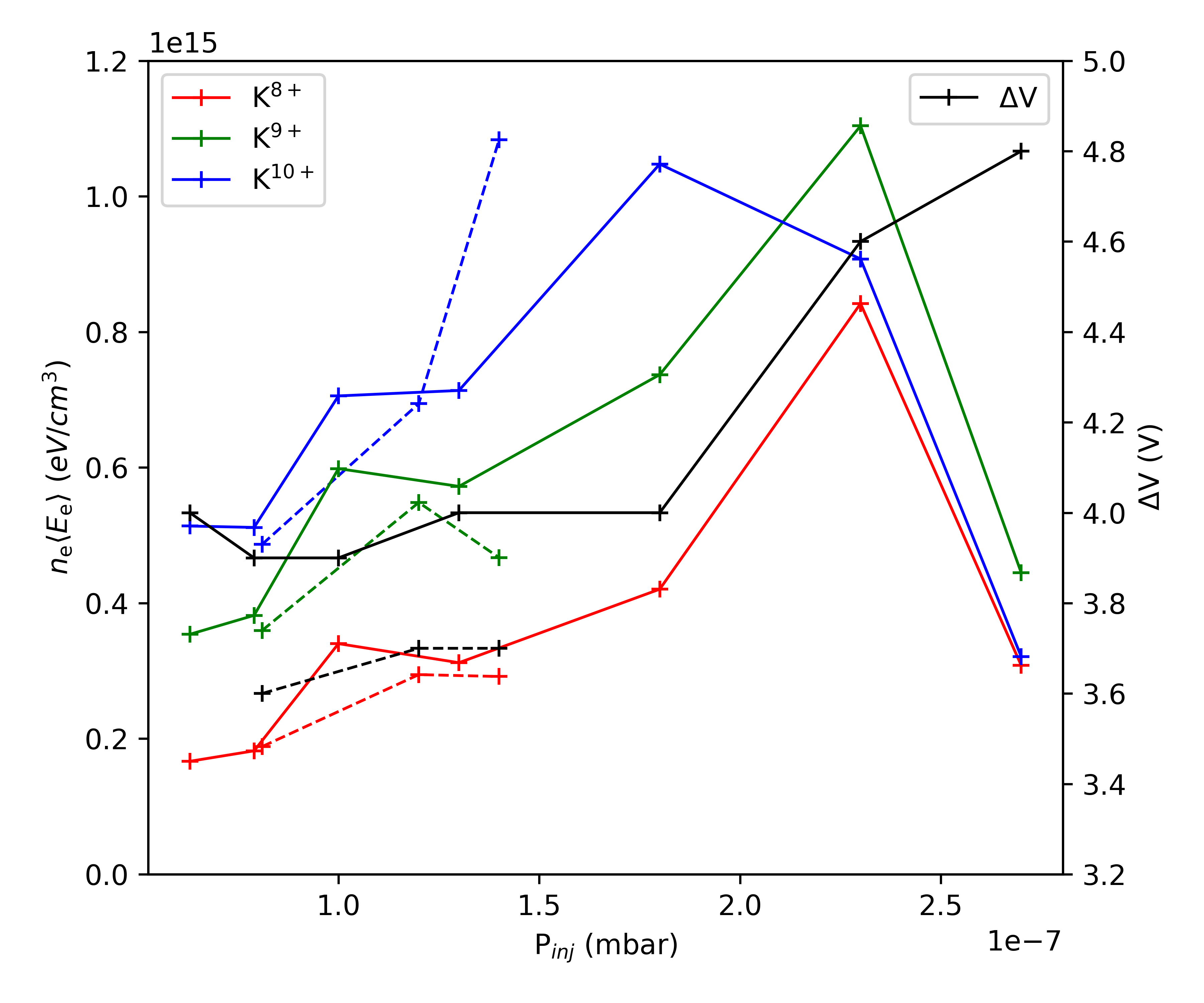}
\caption{\label{fig:gas12_EnergyContent} The (local) plasma energy content of K$^{8+}$, K$^{9+}$ and K$^{10+}$ together with the optimum $\Delta$V value as a function of the H$_2$ (buffer) gas pressure. The solid and dashed lines represent two different pressure sweeps.}
\end{figure}

The effect of the H$_2$ neutral pressure on the plasma energy content is shown in Fig.~\ref{fig:gas12_EnergyContent} for potassium charge states from K$^{8+}$ to K$^{10+}$. Here the results of both sweeps are displayed (either solid or dashed lines). It is seen that the trend of the plasma energy content is to increase with the gas pressure (with the exception of the highest pressure), which is attributed to higher plasma density as indicated by the histograms of the solution sets shown as projections to the axes in Fig.~\ref{fig:solution_set_example}. The finding is commensurate with diamagnetic loop experiments reporting the plasma energy content to increase with the neutral gas pressure (saturating at high pressures)~\cite{Noland}. Figure~\ref{fig:gas12_EnergyContent} also shows the evolution of the $\Delta$V-value. The $\Delta$V appears to follow the same trend as the energy content. This is consistent as the optimum energy (tuned with $\Delta$V) for 1+ beam capture relies on the plasma potential \cite{Tarvainen_stopping}, which presumably depends on the low energy electron density as implied by the data in Ref.~\cite{Tarvainen_plasmapotential}.

\begin{figure*}
\includegraphics[width=\textwidth]{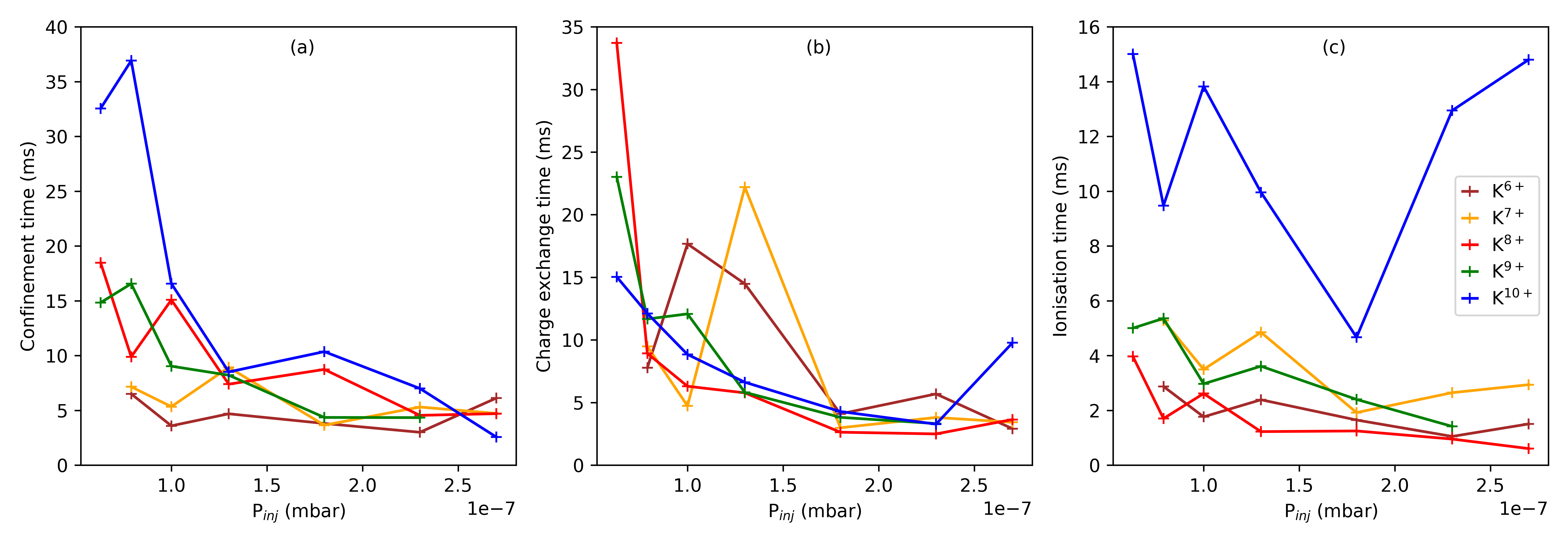}
\caption{\label{fig:gas_Time} The confinement, charge exchange and ionisation times of K$^{6+}$ - K$^{10+}$ as a function of the H$_2$ (buffer) gas pressure.}
\end{figure*}

Figure \ref{fig:gas_Time} shows $\tau_{\textrm{conf}}$, $\tau_{\textrm{cex}}$ and $\tau_{\textrm{ion}}$ for charge states K$^{6+}$-K$^{10+}$ as a function of the H$_2$ neutral gas pressure. All these characteristic times tend to decrease with the neutral (buffer) gas pressure, with only the ionisation time of K$^{10+}$ breaking the trend. We note that obvious outlier points, e.g. $>$\SI{1}{\second} $\tau_{\textrm{cex}}$ for K$^{6+}$ arising from poor fits to experimental transient data are not shown in the figure, which explains 'missing data points' at low pressure.

\subsection{Magnetic field minimum, B$_{\textrm{min}}$}\label{sec:bmin} 

\begin{figure}
\includegraphics[width=0.5\textwidth]{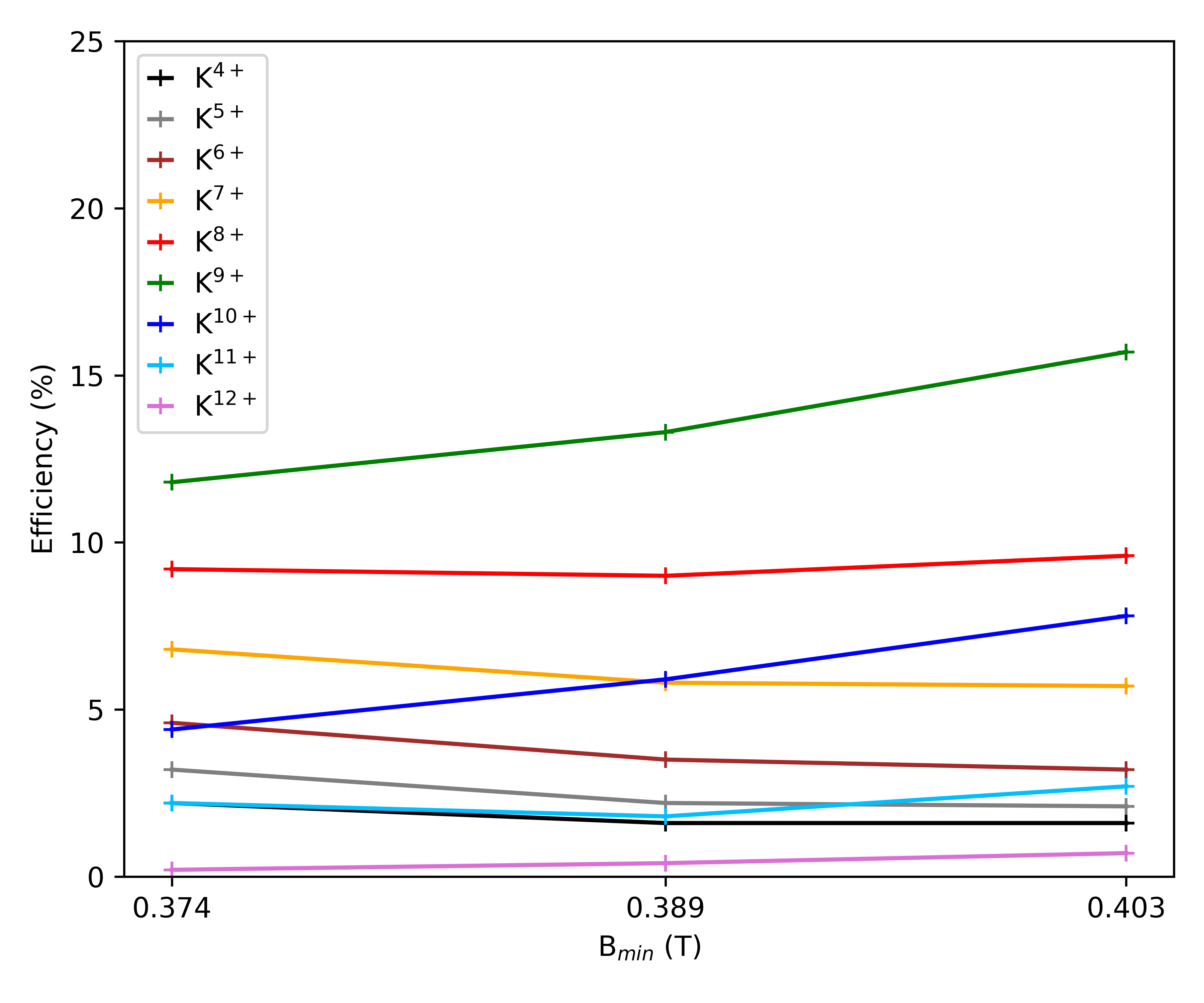}
\caption{\label{fig:BMin_Efficiency} The charge breeding efficiency of K$^{4+}$ - K$^{12+}$ as a function of B$_{\textrm{min}}$.}
\end{figure}

The charge breeding efficiencies of K$^{4+}$-K$^{12+}$ at different B$_{\textrm{min}}$ are shown in Fig.~\ref{fig:BMin_Efficiency}. The other source parameters, i.e. magnetic field maxima, microwave power and H$_2$ gas pressure were kept constant at B$_{\textrm{inj}}$ of \SI{1.51}{\tesla}, B$_{\textrm{ext}}$ of \SI{0.82}{\tesla}, \SI{530}{\watt} and \SI{1.1e-7}{\milli\bar}, respectively. The CB efficiency of charge states $\leq$K$^{7+}$ decreases, and the efficiency of charge states $\geq$K$^{8+}$ increases with increasing B$_{\textrm{min}}$.

\begin{figure}
\includegraphics[width=0.5\textwidth]{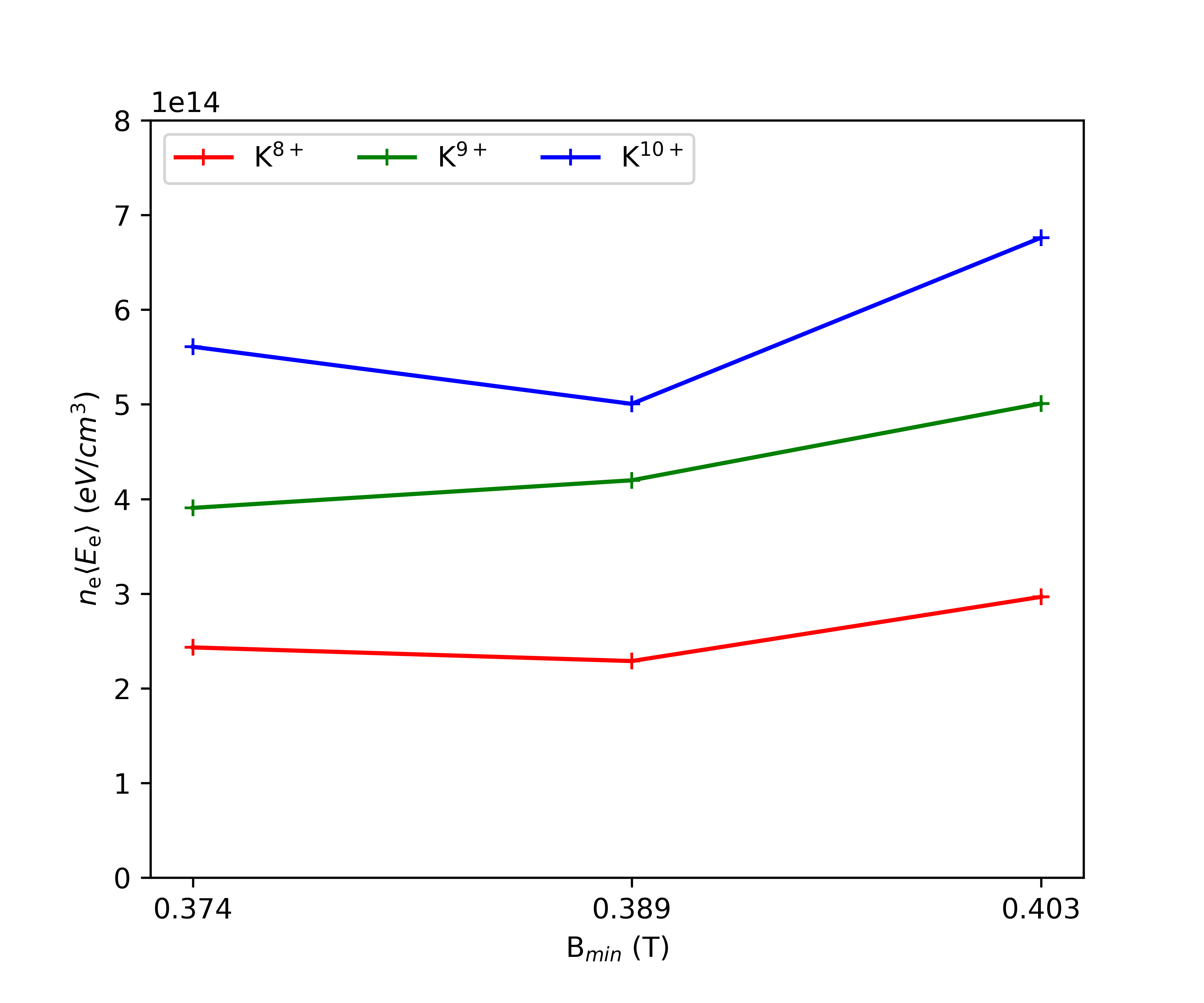}
\caption{\label{fig:BMin_EnergyContent} The (local) plasma energy content of K$^{8+}$, K$^{9+}$ and K$^{10+}$ as a function of B$_{\textrm{min}}$.}
\end{figure}

The plasma energy content at three different B$_{\textrm{min}}$ settings is presented for K$^{8+}$-K$^{10+}$ in Fig.~\ref{fig:BMin_EnergyContent}. The highest energy content is systematically found at the strongest B$_{\textrm{min}}$.

\begin{figure*}
\includegraphics[width=\textwidth]{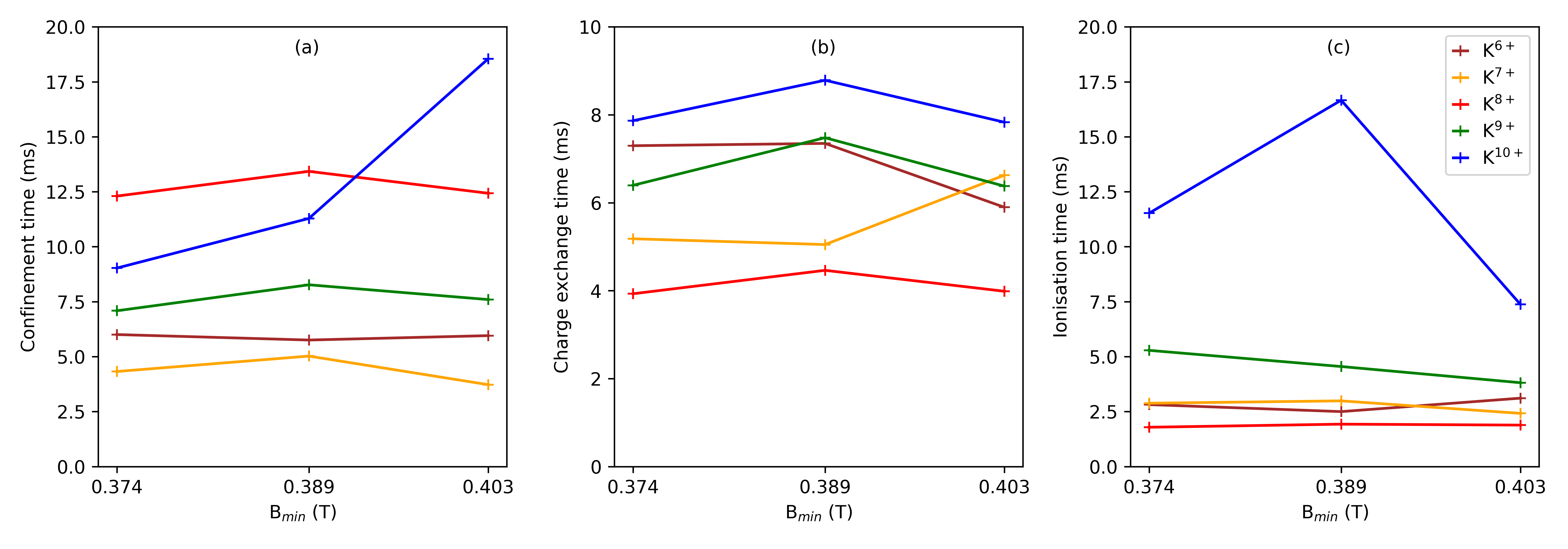}
\caption{\label{fig:BMin_Time} The confinement, charge exchange and ionisation times of K$^{6+}$ - K$^{10+}$ as a function of B$_{\textrm{min}}$.}
\end{figure*}

The characteristic times $\tau_{\textrm{conf}}$, $\tau_{\textrm{cex}}$ and $\tau_{\textrm{ion}}$ for charge states K$^{6+}$-K$^{10+}$ are shown in Fig.~\ref{fig:BMin_Time} as a function of B$_{\textrm{min}}$. There are no systematic trends except the confinement time of the highest K$^{10+}$ charge state approximately doubling from the weakest to strongest B$_{\textrm{min}}$ value, which together with the enhanced CB efficiency of even higher charge states implies improved (electrostatic) ion confinement.

\subsection{Extraction mirror magnetic field, B$_{\textrm{ext}}$}\label{sec:bext} 

\begin{figure}
\includegraphics[width=0.5\textwidth]{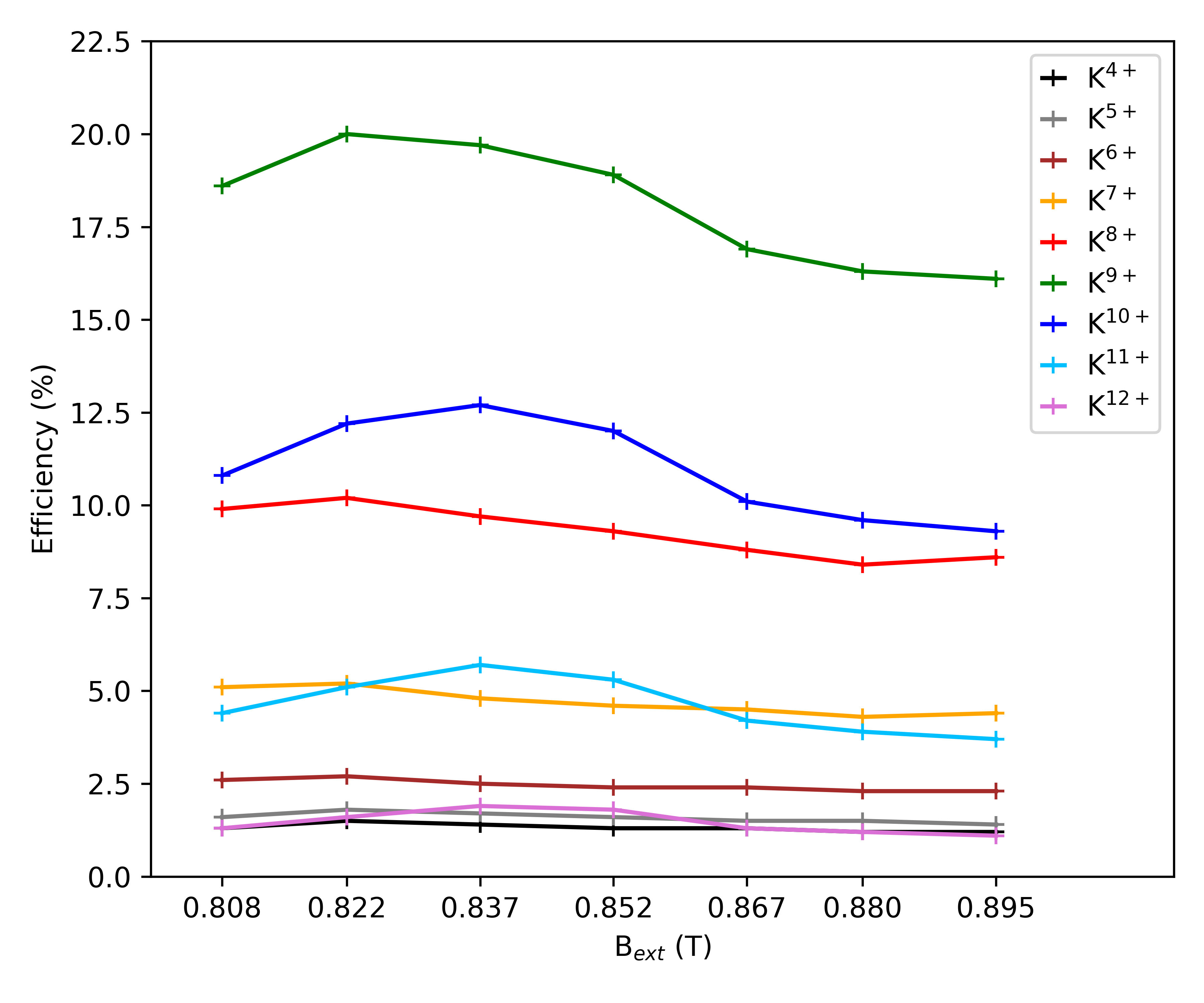}
\caption{\label{fig:Bext_Efficiency} The charge breeding efficiency of K$^{4+}$ - K$^{12+}$ as a function of B$_{\textrm{ext}}$.}
\end{figure}

The effect of the extraction mirror field B$_{\textrm{ext}}$ on the charge breeding efficiencies of K$^{4+}$-K$^{12+}$ is illustrated in Fig.~\ref{fig:Bext_Efficiency}. Here the other source parameters were as follows: B$_{\textrm{inj}}$ of \SI{1.58}{\tesla}, B$_{\textrm{min}}$ of \SI{0.45}{\tesla}, \SI{530}{\watt} microwave power and \SI{1.4e-7}{\milli\bar} H$_2$ pressure. The CB efficiency of high charge state ions, i.e. K$^{9+}$ and higher, exhibits a clear optimum at \SI{0.83}-\SI{0.84}{\tesla} while the efficiency of lower charge states decreases monotonically with increasing B$_{\textrm{ext}}$.

\begin{figure}
\includegraphics[width=0.5\textwidth]{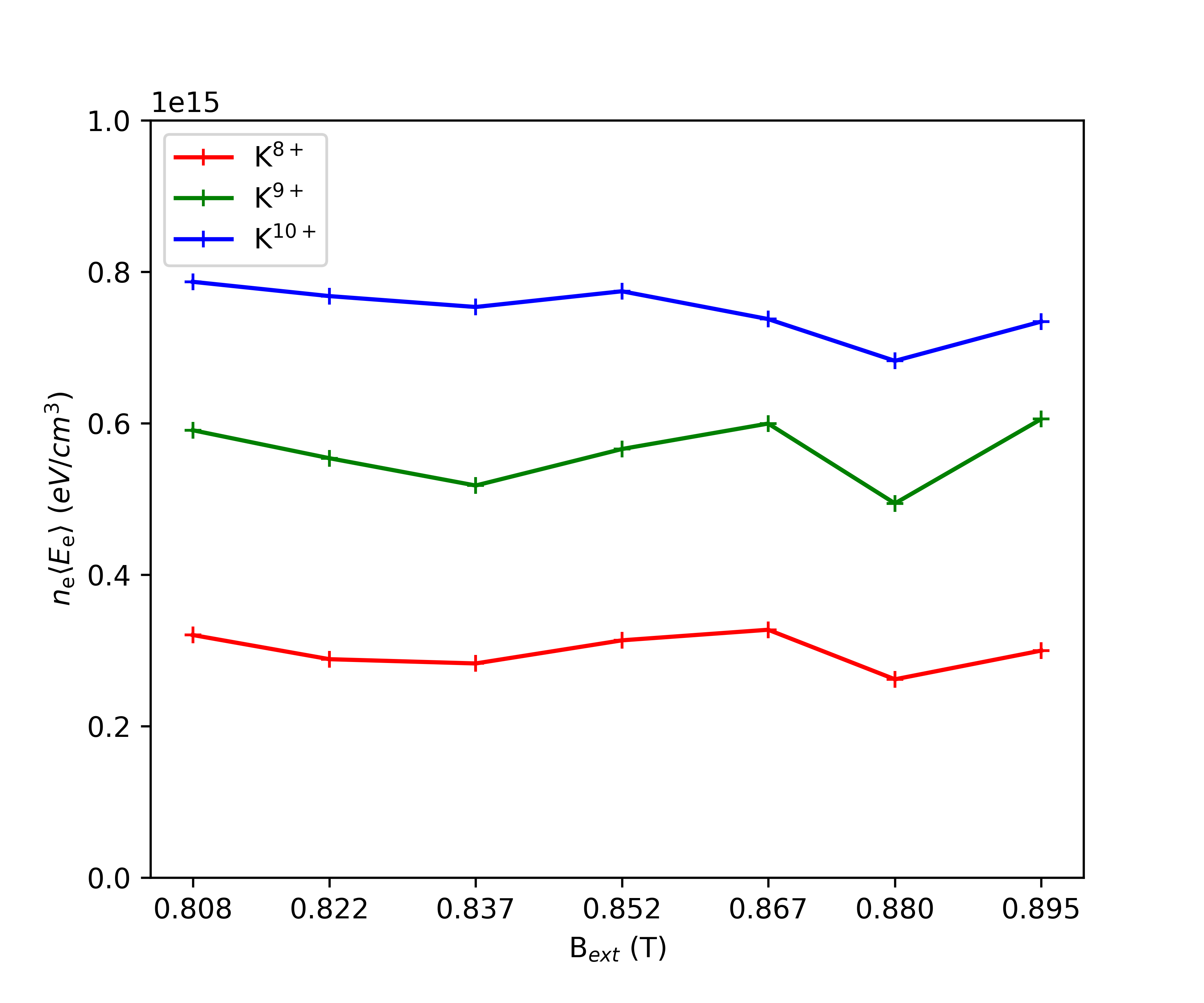}
\caption{\label{fig:Bext_EnergyContent} The (local) plasma energy content of K$^{8+}$, K$^{9+}$ and K$^{10+}$ as a function of B$_{\textrm{ext}}$.}
\end{figure}

Figure \ref{fig:Bext_EnergyContent} shows the plasma energy content (for K$^{8+}$-K$^{10+}$) as a function of B$_{\textrm{ext}}$. The extraction field has very little effect on the energy content, i.e. there is no trend observed with this parameter. 

\begin{figure*}
\includegraphics[width=\textwidth]{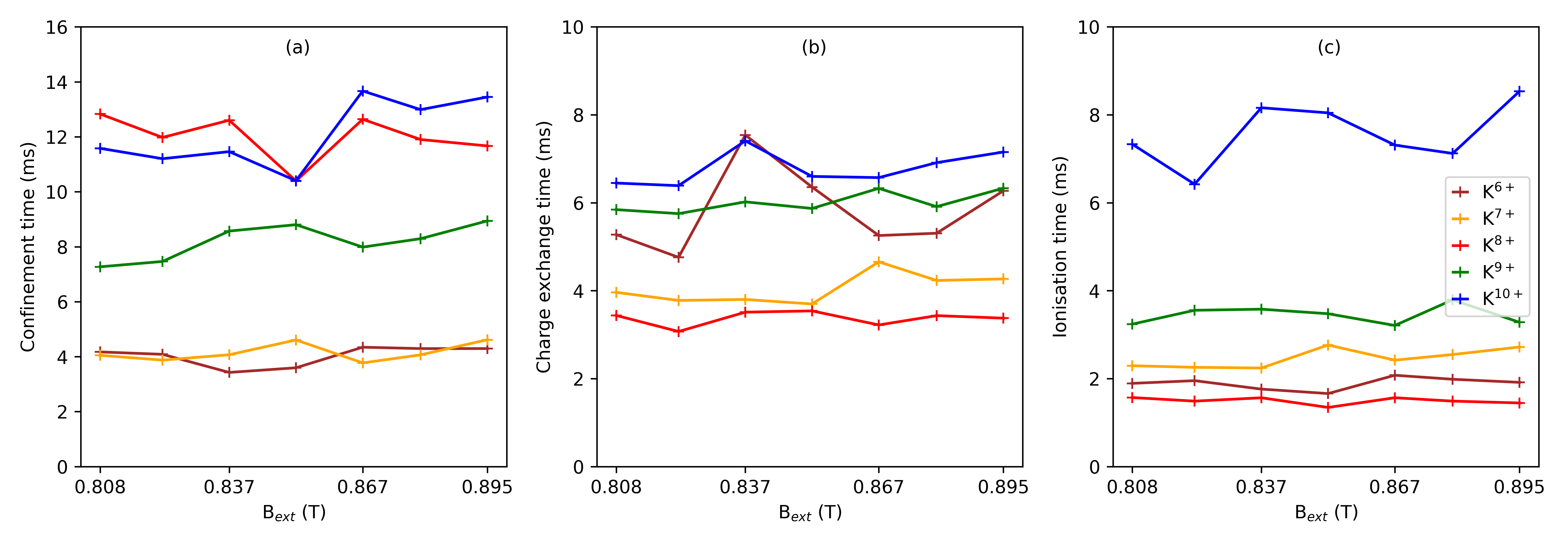}
\caption{\label{fig:Bext_Time} The confinement, charge exchange and ionisation times of K$^{6+}$ - K$^{10+}$ as a function of B$_{\textrm{ext}}$.}
\end{figure*}

The characteristic times $\tau_{\textrm{conf}}$, $\tau_{\textrm{cex}}$ and $\tau_{\textrm{ion}}$ of charge states K$^{6+}$-K$^{10+}$ measured with different B$_{\textrm{ext}}$ are shown in Fig.~\ref{fig:Bext_Time}. There are no clear trends, which is in line with B$_{\textrm{ext}}$ having little effect on the CB efficiency compared to e.g. the neutral gas pressure. Nevertheless, we note that the efficiency increase of the optimum charge state K$^{9+}$ at B$_{\textrm{ext}}$ between \SI{0.822}{\tesla} and \SI{0.852}{\tesla} corresponds to an increase of the confinement time.

\subsection{On the uncertainty of $\tau_{\textrm{conf}}$, $\tau_{\textrm{cex}}$ and $\tau_{\textrm{ion}}$} \label{sec:uncertainty}

As stated earlier, the most prominent downside of using the CT method for estimating the characteristic times $\tau_{\textrm{conf}}$, $\tau_{\textrm{cex}}$ and $\tau_{\textrm{ion}}$ of the high charge state ions is the large uncertainty. Typical uncertainties of high charge state potassium confinement and charge exchange times are 100-200\% while the ionization times can be estimated more accurately, i.e. with 40-70\% relative uncertainty \cite{Angot_CT, Luntinen_ICIS21, Luntinen_PhysRevE, Luntinen_uncertainty}, which raises the concern that the CT method might not be able to detect small variations of the plasma parameters. Thus, the statistical relevance of the measurement results presented above could be questioned. However, it has been shown in Ref.~\cite{Luntinen_uncertainty} that the large uncertainties are inherited from the ionization cross section data \cite{Lennon}. This allows us to argue that the CT method can reveal trends of the characteristic times as a function of a control parameter, such as microwave power, gas pressure and magnetic field strength, although the absolute values of the times are subject to systematic errors of the cross section data. In other words, the conclusions based on the trends of the characteristic time median values displayed in Figs.~\ref{fig:uw_time}, \ref{fig:gas_Time}, \ref{fig:BMin_Time} and \ref{fig:Bext_Time} are not affected by the uncertainties of the individual data points. Hence, the data are presented without the corresponding uncertainties for the clarity of the illustration. 

\section{Conclusions and discussion}
\label{discussion}

\begin{figure}
\includegraphics[width=\columnwidth]{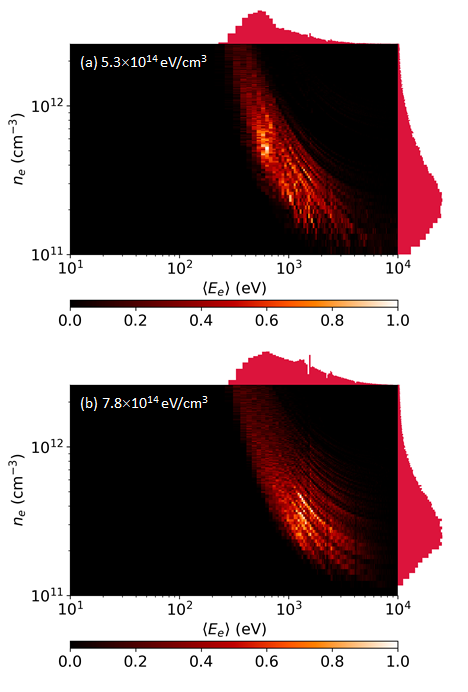}
\caption{\label{fig:solution_set_uw} The $\left ( \left < E_e \right > , n_e \right )$ solution sets for K$^{10+}$ and the corresponding plasma energy contents. The microwave power was increased from \SI{150}{\watt} in (a) to \SI{600}{\watt} in (b). The energy content increases due to increase in the median electron energy from \SI{1.0}{\kilo\electronvolt} to \SI{1.5}{\kilo\electronvolt}.}
\end{figure}

It was found that the charge breeding efficiency of high charge state K ions and the plasma energy content in the core of the ECR discharge increases with the microwave power. Closer inspection of the solution set histograms reveals that this is most likely due to increase of the median $\left < E_e \right >$ rather than $n_e$ as illustrated in Fig.~\ref{fig:solution_set_uw} showing the $\left ( \left < E_e \right > , n_e \right )$ solution sets for K$^{10+}$ at microwave powers of \SI{150}{\watt} (a) and \SI{600}{\watt} (b) as a representative example. This interpretation is commensurate with the shift of the charge state distribution (charge breeding efficiency vs. charge state) as the peak of the ionisation cross section is at higher energy for the high charge ions. The increase of the plasma energy content with the microwave power has been observed earlier with diamagnetic loop diagnostic \cite{Noland}. No sound conclusions can be made from the characteristic times as a function of the microwave power.

Neutral gas pressure was found to be the control parameter producing the clearest trends of the plasma energy content and characteristic times. The increase of the plasma energy content with the neutral gas pressure is attributed to the increase of the plasma density rather than the average energy of the warm electrons (see Fig.~\ref{fig:solution_set_example}). Similar conclusion, i.e. increase of $n_e$ with the neutral gas pressure, has been drawn when probing ECRIS plasmas with diamagnetic loop \cite{Noland}, K-alpha x-ray emission \cite{Sakildien} or 1+ in-flight ionisation in a charge breeder \cite{Emilie}. The behavior of the charge breeding efficiency with the buffer gas pressure can be explained as follows: as the neutral gas feed rate is increased, the enhanced charge exchange rate limits the high charge state ion production whereas ions with low or medium charge state first benefit from the increased plasma density (electron impact ionisation rate) and only at very high neutral gas pressure their production is limited by the charge exchange. This interpretation is supported by the fact that the charge exchange times have a decreasing trend with the pressure. It is worth noting that the ionisation time of the highest charge states is shorter than their charge exchange time only at the lowest pressure, which matches the observed trend of the charge breeding efficiency. The decrease of the confinement time with the pressure is attributed to higher plasma density, which increases the electron flux and ambipolar plasma potential (see 
e.g.~\cite{Tarvainen_plasmapotential}), thus reducing the average ion confinement time as the fluxes of negative and positive charge carriers are equal in equilibrium condition (i.e. $n_e/\tau_{e}=\Sigma q n_i^q/(\tau_{\textrm{conf}}^q)$ where $q$ refers to the charge state of the ion). The decrease of the charge exchange and ionisation times with the neutral gas pressure are presumably due to increased neutral and plasma (electron) densities affecting the charge exchange and electron impact ionisation rates $n_nn_i\left<\sigma_{\textrm{cex}}v_i\right >$ and $n_en_i\left<\sigma_{\textrm{ion}}v_e\right >$, respectively. 

\begin{figure}
\includegraphics[width=\columnwidth]{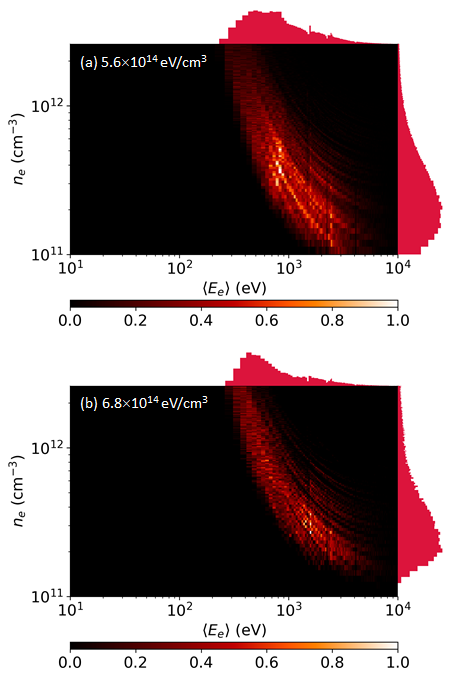}
\caption{\label{fig:solution_set_Bmin} The $\left ( \left < E_e \right > , n_e \right )$ solution sets for K$^{10+}$ and the corresponding plasma energy contents. The $B_{\textrm{min}}$ was increased from \SI{0.37}{\tesla} in (a) to \SI{0.40}{\tesla} in (b). The energy content increases slightly due to increase in the median electron energy from \SI{1.1}{\kilo\electronvolt} to \SI{1.3}{\kilo\electronvolt}.}
\end{figure}

Examining the solution sets obtained in the extremes of the minimum-B magnetic field sweep reveals that the small increase of the plasma energy content with $B_{\textrm{min}}$ is most likely due to increasing $\left < E_e \right >$ rather than $n_e$ as illustrated in Fig.~\ref{fig:solution_set_Bmin} showing the $\left ( \left < E_e \right > , n_e \right )$ solution sets for K$^{10+}$ at $B_{\textrm{min}}$ of \SI{0.37}{\tesla} (a) and \SI{0.40}{\tesla} (b). This is consistent with $B_{\textrm{min}}$ being the most influential parameter affecting the plasma bremsstrahlung spectral temperature \cite{Benitez} and the occurrence of kinetic instabilities \cite{Tarvainen_instability} driven by the anisotropy of the hot electron component \cite{Golubev}. Finally, $B_{\textrm{ext}}$ sweep yielded very similar solution sets (not shown for brevity) and median $n_e$ and $\left < E_e \right >$ -values regardless of the absolute strength of the extraction mirror field, i.e. $B_{\textrm{ext}}$ appears to have a smaller effect on the plasma parameters than $B_{\textrm{min}}$ as discussed in Ref.~\cite{Toivanen_ganil}.

Besides the parametric trends of the plasma energy content and characteristic times we can make some general observations. The conditions for producing fully stripped ions in ECRIS plasma have been postulated in Ref.~\cite{Golovanivsky} presenting the so-called Golovanivsky plot displaying the required product $n_e\tau_{\textrm{conf}}$ at the optimum electron temperature $T_e$ of a Maxwellian distribution with $kT_e= \left < E_e \right > 2/3 $ to produce various fully stripped ions. For argon, which is the neighbouring element to potassium, the triple product $n_e\tau_{\textrm{conf}}T_e$ required for fully stripped ions is approximately $3.2\times 10^{15}$\SI{}{\electronvolt\second/\centi\meter\cubed}. In this work we have found that the plasma energy content $n_e\left < E_e \right >$ ranges from $0.2\times 10^{15}$\SI{}{\electronvolt/\centi\meter\cubed} to $1.0\times 10^{15}$\SI{}{\electronvolt/\centi\meter\cubed} with \SI{10}-\SI{15}{\milli\second} confinement times for the highest charge states of potassium. These values translate to triple product of $0.1$-$1.0\times 10^{13}$\SI{}{\electronvolt\second/\centi\meter\cubed} suggesting that fully stripped K$^{19+}$ ions cannot be produced with the CB-ECRIS, which is commensurate with the extracted charge state distribution where the maximum detectable charge state of potassium is K$^{12+}$.

Further to the absolute scale of the triple product (plasma parameters) we note that the results reveal a hierarchy of characteristic times relevant for high charge state ion production. The peak of the charge breeding efficiency distribution is at K$^{9+}$, which is the highest charge state for which we find consistently $\tau_{\textrm{ion}}<\tau_{\textrm{cex}}$ and $\tau_{\textrm{ion}}<\tau_{\textrm{conf}}$. In other words, for charge states 9+ and lower, the ionisation time is shorter than the charge exchange time or the ion confinement time, which causes ions to "pile-up" on that charge state making its charge breeding efficiency the highest. For charge states above the peak of the breeding efficiency distribution the time hierarchy appears to convert to $\tau_{\textrm{cex}}<\tau_{\textrm{conf}}$ and $\tau_{\textrm{cex}}<\tau_{\textrm{ion}}$, i.e. charge exchange limits the production of very high charge state ions.   

Overall, the results discussed here have demonstrated that despite of the large relative uncertainty (see Ref.~\cite{Luntinen_uncertainty} for thorough discussion) the CT method is sensitive enough to identify trends in the plasma parameters, e.g. the increase of the plasma density with the neutral gas pressure. Importantly, these trends are similar to those inferred from other diagnostics such as diamagnetic loop experiments and K-alpha emission 
\cite{Noland, Sakildien}. The advantage of the CT method over other diagnostics techniques arises from its simplicity; practically all charge breeders are readily equipped with 1+ beam pulsing and N+ beam current (transient) detection apparatus. The computational analysis tools required to translate the beam current transients into $\left ( \left < E_e \right > , n_e \right )$ solution sets and corresponding characteristic times, $\tau_{\textrm{conf}}$, $\tau_{\textrm{cex}}$ and $\tau_{\textrm{ion}}$, are open source and available through GitHub \cite{github}.



\begin{acknowledgments}
We acknowledge grants of computer capacity from the Finnish Grid and Cloud Infrastructure (persistent identifier urn:nbn:fi:research-infras-2016072533), and support of the Academy of Finland Project funding (Grant No:315855).
\end{acknowledgments}



\end{document}